# ENTROPY BOUND FOR THE CRYSTALLINE VACUUM
# COSMIC SPACE MODEL

J. A. Montemayor-Aldrete[1], J. R. Morones-Ibarra[2], A. Morales-Mori[3],

A. Mendoza-Allende[1], E. Cabrera-Bravo and A. Montemayor-Varela[4].

1   Instituto de Física, Universidad Nacional Autónoma de México,  Apartado Postal 20-364, 01000 México, D. F.

2   Facultad de Ciencias Físico - Matemáticas, Posgrado, Universidad Autónoma de Nuevo León.  Apartado Postal 101 - F, 66450 San Nicolás de los Garza N. L. México.

3   Centro de Ciencias Físicas, Universidad Nacional Autónoma de México, Apartado Postal 139-B, 62191 Cuernavaca, Morelos México.

4   Centro de Mantenimiento, Diagnóstico y Operación, Iberdrola, Polígono Industrial El Serrallo, 12100, Castellón de la Plana, España.

**ABSTRACT**

By applying the Heisenberg's uncertainty principle for a macroscopic quantum gas formed by gravitational waves an expression for the universal bound on the entropy proposed by Bekenstein for any system of maximum radius $R$ and total energy $E$ has been obtained. By using such expression, in the theoretical scheme of the crystalline vacuum cosmic model, the low entropy value at the Big Bang beginning, $10^{88}k$, is explained. According to our analysis the time arrow is well defined and the theoretical time flow occurs only in one direction as requested by the physical processes of gravitational stabilization of the vacuum space crystalline structure around equilibrium conditions.

# I.    INTRODUCTION

In 1981, Bekenstein [1] proposed the existence of a universal bound on the entropy $S$ of any object or system of maximum radius $R$ and total energy $E$,

$$S_{max} \leq \frac{2\pi k\, RE}{\eta c} \tag{1}$$

Where $\eta, k$ and $c$ have the usual meaning. According to Bekenstein [2], this bound was inferred from the requirement that the generalized second law of thermodynamics for black hole [3 -6 ] be respected when a box containing entropy is deposited with no radial motion next to the horizon of a Schuarzschild black hole, and then allowed to fall in. The box's entropy disappears but an increase in the black hole entropy occurs. The second law is respected provide $S$ is bounded as in Eq. (1). An analytical proof of the bound has been obtained by Schiffer and Bekenstein [7] for some generic system consisting of a non interacting quantum field in three dimensions confined to a cavity of arbitrary shape and topology.

The main goal of this paper is twofold. First, to obtain an expression similar to Eq. (1) starting from the Heisenberg's uncertainty principle. And second, to apply the obtained equation for the crystalline vacuum cosmic space model to describe the change in entropy before and after the matter packages formation occurring in the early Universe.

## 2. THEORY

For the sake of making this paper self contained, let us make a brief synthesis of some previous results obtained from the crystalline vacuum cosmic model [8 - 11] which will be used in our analysis.

From the gravitational stability analysis of a crystalline vacuum cosmic space [9], with lattice parameter of the order of the neutron radius, an expression which describes the adiabatic compression of the original gravitational waves for the gravitational stability of crystalline vacuum cosmic space has been obtained, namely

$$\frac{T(r)}{T_{OU}(R_{OU})} = \frac{R_{OU}}{r}, \qquad (2)$$

with $R_{OU}$ as the present Universe radius and $T_{OU}(R_{OU})$ the absolute temperature associated to the original stabilization gravitational waves with energy $\varepsilon_{OU}$, given by,

$$\varepsilon_{OU} \equiv \Delta\varepsilon_{OU} \geq \frac{\eta c}{R_{OU}} = k\,T_{OU} \qquad (3)$$

and $T(r)$ as the gravitational wave temperature with concentric radius $r$. In this scheme $R_{OU}$ encloses $10^{120}$ physical lattice points, and the gravitational stabilization of such crystalline region is obtained by forming $10^{80}$ baryons (in particular neutrons) starting from $10^{120}$ original gravitational stabilization waves each one described by Eq. (3), [9]. In other words, the collective long - range quantum fluctuation which stabilizes the vacuum crystalline structure against gravitational instabilities has an energy density, $U_{qf}$, of about $10^{-40}$ times the energy density of the crystalline vacuum structure, $U_{VS}$, [9, 11]. In terms of internal energy of the crystalline structure of the vacuum space, $E_{VS}$, the total change in the crystal internal energy, inside the volume $V_{OU} = \frac{4}{3}\pi R_{OU}^3$, due to the long - range quantum fluctuation, $E_{qf}$, is $E_{qf} = 10^{-40} E_{VS}$.

Also, another important result obtained previously [9], from this adiabatic compression, by its own self attraction, of the original gravitational waves for the stabilization of the crystalline vacuum cosmic space is,

$$T(r)\ \Delta t(r) \geq \eta / k,$$
(4)

with $\Delta t(r) = r/c.$

At first approach, Eq. (4) gives an apparent paradoxical situation: the temperature measurements of a system require a time interval to be determined when equilibrium thermodynamics, which defines temperature, excludes time for such situation. In our case by using similar expressions as Eq. (3) to describe gravitational transitions, it is easy to obtain $\Delta E \Delta t \geq \eta$ from Eq. (4) which is the Heisenberg's uncertainty principle. In other words temperature, which is a thermodynamical concept used to describe disordered of degraded energy, requires as any other form of energy a finite minimum interval of time to be measured in a quantum scheme $\Delta t \geq \eta / kT$ . From Eq. (4) it is clear that the minimum time required to measure a given temperature only is measurable or important when $T$ is a very low quantity, and for every day purpose it is irrelevant.

Heisenberg's uncertainty principle in the form,

$$\Delta P\ \Delta X \geq \eta,$$
(5)

where $\Delta P$ is the uncertainty in the system linear momentum and $\Delta X$ is the uncertainty in its position will be used to analyze a macroscopic quantum gas as follows. Let us consider a macroscopic quantum gas of volume $V$ on which a force is exerted by an impulsive interaction given by $f \equiv \Delta P / \Delta t$ . Then inequality Eq. (5) could be written as $(\Delta P / A) \Delta V \geq \eta$, where $A$ is for area and $\Delta V$ is the change in volume. Another way to write the previous equation is,

$$\frac{1}{A} \left( \frac{\Delta P}{\Delta t} \right) \Delta V \Delta t \geq \eta \qquad (6)$$

or

$$\{ \Delta \sigma \Delta V \} \Delta t \geq \eta \qquad (7)$$

where $\Delta \sigma \equiv \frac{1}{A} \frac{\Delta P}{\Delta t}$. Also it is clear that for any physical system $\Delta \sigma$ and $\Delta V$ have opposite sign. From the theory of thermodynamic fluctuations [12] we know that, in our notation, $<\Delta \sigma \Delta V> = -kT$. Therefore, Eq. (7) becomes,

$$T \, \Delta t \geq \eta / k \qquad (8)$$

where $T$ is the absolute temperature of the quantum gas. This result is identical to Eq. (4), and the theoretical way used here, enlighten their physical meaning.

During the gravitational stabilizing process of the crystalline vacuum cosmic space we know that the collective long - range quantum fluctuations responsible for such stabilization change the internal energy of the vacuum crystalline space by a quantity $E_{qf}$ (inside a volume $V_{OU}$). Thus by considering that the gravitational waves which cause such collective long - range quantum fluctuation could be considered at first approximation as a macroscopic quantum gas under an adiabatic process of compression due to its own self - attraction, it is possible to obtain an expression for the change in entropy of such a gas a function of the radius $r$, $\Delta S(r)$ as given by

$$\Delta S(r) = \frac{E_{qf}}{T(r)} \qquad (9)$$

and by using Eqs. (8) and (9) and considering that for gravitational waves $R = c\,\Delta t(R)$ it is easy to obtain the following expression,

$$\Delta S(R) \leq \frac{k\,ER}{\eta c},\qquad\qquad\qquad (10)$$

with $E = E_{qf}$, and neglecting a factor of $2\pi$, this equation is equal to Eq. (1) proposed by Bekenstein [1]

Therefore Eq. (10) could be applied to analyze the corresponding situation to the initial instability of the crystalline vacuum cosmic space and the previous state where gravitational waves (from the gravitational stabilization of the crystalline vacuum space) becomes $10^{80}$ quanta matters (neutrons). By using this equation for $R = R_{OU}$ and for $R = r_N$, where $r_N$ is the neutron radius, together with Eq. (2) the following result is obtained for stabilizing gravitational waves,

$$S_{NF} \leq 10^{-40}\,S_{OU}\qquad\qquad\qquad (11)$$

where $S_{OU}$ is the entropy (inside of a Volume $V_{OU}$) at the maximum situation of gravitational instability of the vacuum cosmic space, $S_{NF}$ is the entropy for a situation close to the limit where the energy of each $10^{40}$ original stabilization gravitational waves produces a neutron (each one with $\left(\varepsilon_{OU} = \dfrac{\eta c}{R_{OU}} = k\,T_{OU}\right)$. It is obvious that both entropy levels are referred to the perfect crystal entropy. Equation (11) is another way to say that $T_{NF} \geq 10^{40}\,T_{OU}$, which makes sense because $T_{OF} \sim 1.27*10^{-27}K$, and $T_{NF} \sim 10^{13}K$; this last result can be easily verified.

In our theoretical scheme, each neutron of energy $u_N = 10^{40} h \nu_{OU}$ is the final product of an adiabatic collective gravitational process occurring among $10^{120}$ fundamental gravitons (each one of which has an energy $h \nu_{OU}$). Each fundamental graviton could be considered the smallest unit of curvature that would be allowed according to Heisenberg's uncertainty principle when is applied to our cosmological model of vacuum cosmic space $E_{OU} = h \nu_{OU} = \dfrac{\eta c}{R_{OU}}$. Also it is clear that each neutron could be considered as the greatest unit of curvature that would exist at equilibrium conditions in the crystalline vacuum space; this unit can be called the matter unit of gravitons.

And by taking into account the De Broglie ondulatory scheme as used previously [9], we know that every fundamental graviton requires fulfilling the physical equation,

$$\lambda(r) = 2\pi r \tag{12}$$

and during the adiabatic compression process due to the self - gravitational attraction each of them obeys the following equation,

$$u_N = N(r) h \nu(r) = constant \tag{13}$$

where the neutron is considered at rest. This equation describes the permitted transitions occurring between the orbiting gravitational waves during the adiabatic process of diminishing the radius.

By using Eq. (12) in Eq. (13), $N(r) = 10^{40} \dfrac{2\pi r}{c} \nu_{OU}$; which for $r = R_{OU}$ takes the value $N(R_{OU}) = 10^{40}$ and for $r = r_N$, $N(r_N) = 1$. Therefore, by using Eq. (10) for each of these two cases it is easy to obtain the following result

$$S(R_{OU}) = \frac{k}{\eta c} 10^{40} h \nu_{OU} R_{OU} = 10^{40} k \tag{14}$$

and,

$$S(r_N) = \frac{k}{\eta c} h \nu_N r_N = 1k \tag{15}$$

with $U_N = h \nu_N = 10^{40} h \nu_{OU}$.

The physical meaning of the last results is the following. The decrease in entropy between the thermodynamical equilibrium state corresponding to the greatest gravitational instability on the crystalline structure of the cosmic vacuum, $S_{OU} = 10^{120}k$, and the entropy of the state previous to the formation of $10^{80}$ matter packages (neutrons), which is $S_{NF} = 10^{80}k$, has its origin in the ondulatory character of the gravitational waves which is a quantum property of the energy in the Universe.

## 3. DISCUSSION AND CONCLUSIONS

By using Heisenberg's uncertainty principle to study the fluctuations on a macroscopic quantum gas composed by gravitational waves, an expression proposed by Bekenstein, for a universal bound on the entropy, $S$, of any object or system of maximum radius $R$ and total energy $E$ has been obtained. With the use of such expression in the scheme of the crystalline vacuum cosmic space some physical results has been obtained which allow us to give some discussion and conclusions about it.

1  According to our results, inside the present Universe volume, at the beginning of the cosmological fluctuation cycle which we are analyzing $10^{120}$ original gravitational waves, each of one having energy $h \nu_{OU} = \eta c / R_{OU}$, contribute to stabilize the crystalline structure of the vacuum cosmic space against the long - range gravitational stresses. To this original gravitational waves corresponds an entropy $S_{OU} = 10^{120}k$.    During the adiabatic compression process occurring between these waves by their mutual gravitational interaction eventually (before the Big Bang event (occurred), $10^{80}$ neutrons are formed. Such neutrons have total entropy $S_{NF}$ of $10^{80}k$. Also we know from reference [9] that the diminishing in the gravitational energy of the fundamental gravitational waves, which stabilizes the vacuum cosmic structure, when an a adiabatic self-compression process occur between these waves from radius $R_{OU}$ to the radius which envelopes the $10^{80}$ neutrons formed at the end of the contraction phase is the required energy source which produces an electromagnetic radiation field with total energy of $10^5 E_{ou}$. This energy gives entropy of $10^5 k$ per baryon. There according to our theoretical scheme the starting entropy at the beginning of the big bang is $10^{85}k$. This value is close enough to the total entropy at the big bang an estimated by Penrose [15]. Then the decrease in entropy during this part of the gravitational stabilization cycle on the crystalline structure of the vacuum cosmic space is due to the quantum properties of the gravitational waves.

The radiation energy required for the expansion cycle will be analyzed in a future paper. This is because here our main interest is to explore the entropy implications of our model. It is possible that the energy yield from the transformation process of one quarter of the original baryon matter, neutrons, into helium (about one percent of $E_{OU}$) contribute to increase the Universe entropy in a significant way; $10^8$ photons per baryon as considered by Hoyle, Burbidge and Norlikar [14] but their analysis is a controversial one and do not gives a physical explanation for the dynamics of the expansion cycle. And also our result for the entropy before the beginning of the adiabatic compression process, between gravitational waves for the gravitational stabilizing of the crystalline vacuum cosmic space, namely $10^{120}k$ is in close agreement with the final entropy for the Universe considered a closed one $10^{123}k$. These last results have been obtained by using the Bekenstein - Hawking formula as though the whole Universe had formed a black hole [16].

At this point, we do not need to worry about the full thermodynamical implications of our analysis because we are facing a fluctuation process of the long - range gravitational stresses on the crystalline structure of the vacuum cosmic space around zero gravitational stresses. Consider for instance that the total change in the internal energy of the crystalline structure due to the long - range quantum fluctuations is about $10^{-40}$ the internal energy of the perfect crystalline structure of the cosmic vacuum. And, also we know that the entropy of a closed thermodynamical system can decrease spontaneously as a result of statistical fluctuations [4].

2) In our scheme ([9, 10, 11] and this paper), the arrow of time is well defined and the flow of theoretical time occurs only in one direction as requested by the physical process of gravitational stabilization of the crystalline vacuum cosmic space. This flow of time is in agreement to our physiological and physicological sensation of the progression of time in the daily life. In our cosmological scheme, inside each volume $V_{OU} = \frac{4}{3}\pi R_{OU}^3$, there exist cosmological causes which precede their associated cosmological effects as implicitly are requested by Penrose [18]

3) The probability for our scheme to occur is one; provided that the vacuum cosmic space has a crystalline structure and a lattice parameter of the order of the neutron radius. This probability is a very high number as compared with the probability of spontaneous occurrence for the beginning of the Big Bang standard model: one part in $((10)^{10})^{123}$, which is the figure calculated by Penrose [18].

4) By using the expression for the Schwarzschild radius $R_S = 2GM/c^2$ with $G$ as the gravitational constant, and $M$ for the mass of a black - hole [19] it is clear that expression (1) becomes the Bekenstein - Hawking formula [4, 20] for the black - hole entropy $S_{BH}$,

$$S_{bh} = \frac{4\pi R_S^2}{4}\left(\frac{kc^3}{G\eta}\right) \quad , \tag{16}$$

which, shows that black holes formed by many baryons have entropy proportional to its surface area (measured at $R_S$). Our analysis shows that Eq. (1), due to Bekenstein, when used for the gravitational waves which stabilizes the crystalline structure of vacuum cosmic space, gives an expression for entropy which depends only on $R$ not on $R^2$. This situation and the well known fact that baryons (or neutrons), as the black holes, only exhibit mass, charge and angular momentum as measurable external parameters suggest that two kinds of black holes exist. One kind formed by many baryons [19-32], and another formed by gravitational waves; in future papers both kind of black hole will be analyzed in relation to our theoretical scheme of a crystalline vacuum cosmic space.

**ACKNOWLEDGEMENTS**

We want to specially thank Prof. M. López de Haro for many years of deep discussions and arguments, for his contribution to final shaping of the ideas and for his encouragement not to give up and unorthodox approach to cosmological problems and also we acknowledge to the librarian Technician G. Moreno for her stupendous work and to O. N. Rodríguez Peña for her patient and skilful typing work.